\documentclass[12pt]{iopart}
\usepackage{iopams}
\usepackage{graphicx}

\begin{document}
\title[Competitive noises and quantum interference]{Competitive stochastic noises in coherently driven two-level atoms and quantum interference}
\author{P. Adam\dag, \ddag,\ A. Karpati\dag, \ddag\ , W. Gawlik\S, 
and J. Janszky\dag, \ddag}
\address{\dag\ Department of Nonlinear and Quantum Optics,
    Research Institute for Solid State Physics and Optics,
    Hungarian Academy of Sciences,\\
    P.O. Box 49, H-1525 Budapest, Hungary}
\address{\ddag\ Institute of Physics, 
    University of P\'ecs,\\ 
    Ifj\'us\'ag \'ut 6. H-7624 P\'ecs, Hungary}
\address{\S\  Marian Smoluchowski Physical Institute, \\
Jagellonian University, Reymonta 4, 30-059 Krak\'ow, Poland}
\date{\today}
\begin{abstract}
  A system of coherently-driven two-level atoms is analyzed in
  presence of two independent stochastic perturbations: one due to
  collisions and a second one due to phase fluctuations of the driving
  field. The behaviour of the quantum interference induced by the
  collisional noise is considered in detail. The quantum-trajectory
  method is utilized to reveal the phase correlations between the
  dressed states involved in the interfering transition channels. It
  is shown that the quantum interference induced by the collisional
  noise is remarkably robust against phase noise.  This effect is due
  to the fact that the phase noise, similarly to collisions,
  stabilizes the phase-difference between the dressed states.

\end{abstract}

\submitto{\JOB}

\maketitle

\section{Introduction}

Quantum interference is one of the most intriguing phenomena of
quantum mechanics. Over the past decade several effects in atom-light
interaction have been predicted and demonstrated experimentally, which
have their origin in quantum interference~\cite{Arimondo96,ScullyZ98}.
Some characteristic examples are reduction and cancellation of
absorption~\cite{CardimonaRS82,Imamoglu89,ScullyZG89,Harris89,MandelK93,ToorZZ95,ZhouS97a}
and spontaneous emission~\cite{ZhuCL95,ZhuNS95,XiaYZ96,ZhuS96}, and
narrow resonances in fluorescence~\cite{ZhouS96,ZhouS97,Alzetta}.

A prerequisite of quantum interference between the transition channels is the
existence of some stable time correlation of the atomic system under
consideration.  This time correlation is usually achieved by coherent coupling
of atomic levels in multilevel systems. Noises on the other hand usually
destroy all correlations in the system. Under special circumstances, however,
even the incoherent perturbation can be responsible for the quantum
interference~\cite{PriorBDB81,WilsongordonF83,grynberg}. In a recent
paper~\cite{KarpatiAGLJ02}, a system of two-level atoms strongly driven by a
coherent light field and perturbed by collisional noise was analyzed by the
quantum-trajectory method~\cite{carmichael-book,PlenioK98,GarrawayK94,DalibardCM92,GardinerPZ92,DumZR92}. It was shown that collisional noise can
unexpectedly create phase correlation between the neighboring atomic dressed
states. This phase correlation is responsible for quantum interference between
dressed state transition channels, resulting in anomalous modifications of the
resonance fluorescence spectra found experimentally earlier~\cite{gawlik}.

Another typical example of a stochastic perturbation of a system of coherently
driven two-level atoms is the phase noise of the laser field.  The effect of
the linewidth of the driving laser on the spectrum has been widely
investigated in the
literature~\cite{Agarwal76,Eberly76,KimbleM77,KnightMS78,ToorZ94}, but no
quantum interference effects have been found. In reference~\cite{ZhouFZL99},
however, it is suggested that an interference phenomenon similar to that
induced by collisional noise can appear due to a large phase noise of the
driving laser field. 
The consideration presented there is valid only in the case when the
correlation between the atomic density matrix and the phase noise can be
neglected. The assumption is rather unphysical and it is not valid, for
instance, for Gaussian noise.

In this paper we present the detailed analysis of a system of coherently
driven two level atoms, when both collisional and phase noises are present.
We carry out the complete quantum-trajectory simulation of the system, which
reveals the underlying physical processes in quantum interference phenomena.

The paper is organized as follows. In section~\ref{sec:thesystem} the model
describing the system is introduced, and the master equation is constructed
for the atomic density operator as a set of ordinary differential equations.
In section~\ref{sec:thespectrum} the resonance fluorescence spectrum is
obtained. In section~\ref{sec:phase} the quantum-trajectory analysis of the
problem is carried out. Based on this, quantum interference effects are
discussed in the case when both collisional and phase noises are present.
Section~\ref{sec:sum} summarizes our results.

\section{The model}
\label{sec:thesystem}

Let us consider a system of two-level atoms driven by a coherent laser
field, incoherently perturbed by elastic, dephasing collisions of the
atoms.  Furthermore let us assume, that the driving laser field has a
finite bandwidth due to its phase noise. We use the following
notation: $\omega_a(t)$ is the fluctuating atomic transition frequency
describing the stochastic perturbation due to collisions, $\varphi(t)$
is the randomly varying laser phase, $\omega_L$ is the frequency of
the laser, $\Omega$ is the Rabi frequency, while $S^z$, $S^+$, $S^-$
are the atomic operators defined on the ground state ($|g\rangle$) --
excited state ($|e\rangle$) basis:
\begin{equation}
S^z=\left(\begin{array}{cc}1/2&0\\0&-1/2\end{array}\right),\ 
S^+=\left(\begin{array}{cc}0&1\\0&0\end{array}\right),\ 
S^-=\left(\begin{array}{cc}0&0\\1&0\end{array}\right).
\end{equation}
The functions $\omega_a(t)$ and $\varphi(t)$ are supposed to describe
Gaussian noise:
\begin{equation}
\omega_a(t) = \omega_a + \delta\omega_a(t),
\end{equation}
\begin{equation}
\langle \delta\omega_a(t)\delta\omega_a(t')\rangle = 2\Gamma\delta(t-t'),
\end{equation}
\begin{equation}
{\rm d\over {\rm d} t}\varphi(t) = \vartheta(t),
\end{equation}
\begin{equation}
\langle \vartheta(t)\vartheta(t')\rangle = 2L\delta(t-t'),
\end{equation}
where $\Gamma$ and $L$ stand for the magnitude of the corresponding
stochastic noises. The phase noise and the noise in the atomic
resonance frequency are independent,
$\langle\delta\omega_a(t)\vartheta(t')\rangle = 0$. Under such
circumstances, the atomic Hamiltonian, in the interaction picture and
rotating wave approximation, is of the form
\begin{equation}
H=\hbar (\omega_a(t)-\omega_L) S^z + {1\over 2}\hbar \Omega(e^{-i\varphi(t)}S^-+S^+e^{i\varphi(t)}).
\label{eq:hamiltonian}
\end{equation}
We remark, that the interaction picture is defined via the
noiseless interaction-free Hamiltonian
\begin{equation}
  \label{eq:hnull}
  H_0= \hbar \omega_a^{(g)} |g\rangle\langle g| +
\hbar \omega_a^{(e)} |e\rangle\langle e|,
\end{equation}
where $\omega_a^{(g)}$ and $\omega_a^{(e)}$ are the original (thus the
mean value) frequencies of the ground state and excited state of the
atom. The noise due to collision is taken into account in the
interaction part of the Hamiltonian.

The master equation describing the time evolution of the system is of the form
\begin{equation}
\dot\rho={1\over i\hbar}[H,\rho] + ({\mathcal{L}}\rho)_{\rm sp},
\label{eq:master}
\end{equation}
where the linear term 
\begin{equation}
({\mathcal{L}}\rho)_{\rm sp}=\gamma(-{1\over 2} (S^+S^-\rho + \rho S^+
S^-)+S^-\rho S^+)
\end{equation}
describes the spontaneous emission and $\gamma$ is the natural linewidth.

The master equation (\ref{eq:master}) is equivalent to the following
stochastical differential equations for the matrix elements of the density
operator, written on the ground state ($|g\rangle$) -- excited state
($|e\rangle$) basis:
\begin{eqnarray}
\dot{\rho}_{ee}&=&{1\over 2}i\Omega (e^{i\varphi(t)}\rho_{ge} - e^{-i\varphi(t)}\rho_{eg})-2\gamma\rho_{ee},\\
\dot{\rho}_{eg}&=&i(\Delta+\delta\omega(t))\rho_{eg} - {1\over 2}i\Omega e^{i\varphi(t)}(\rho_{ee}-\rho_{gg})\nonumber\\&&-\gamma\rho_{eg},\\
\dot{\rho}_{ge}&=&-i(\Delta+\delta\omega(t))\rho_{ge} + {1\over 2}i\Omega e^{-i\varphi(t)}(\rho_{ee}-\rho_{gg})\nonumber\\&&-\gamma\rho_{ge},\\
\dot{\rho}_{gg}&=&-{1\over 2}i\Omega (e^{i\varphi(t)}\rho_{ge} - e^{-i\varphi(t)}\rho_{eg})+2\gamma\rho_{ee},
\end{eqnarray}
where $\Delta = \omega_a - \omega_L$ is the detuning of the laser from the
atomic resonance frequency. The stochastic differential equations can be
reduced  to ordinary differential equations by using the theory of
multiplicative stochastic processes~\cite{Agarwal78}. We summarize the main
steps of this method briefly. The first step is to introduce new variables in
the form $\chi_1 =
\rho_{eg},\,\chi_2=\rho_{ge}e^{2i\varphi},\,\chi_3=\rho_{gg}e^{i\varphi},\,\chi_4=\rho_{ee}e^{i\varphi}$.
The differential equation system in these new variables is linear and does not
contain the variable $\varphi(t)$ explicitly, only its time-derivative
$\vartheta(t)$:
\begin{equation}
\dot{\chi_i} = \sum_j M_{ij}\chi_j + F_{ij}(t)\chi_j,
\end{equation}
where
\begin{equation}
M=\left[\begin{array}{cccc}
i\Delta-\gamma&0&{1\over 2}i\Omega&-{1\over 2}i\Omega\\
0&-i\Delta-\gamma&-{1\over 2}i\Omega&{1\over 2}i\Omega\\
{1\over 2}i\Omega&-{1\over 2}i\Omega&0&2\gamma\\
-{1\over 2}i\Omega&{1\over 2}i\Omega&0&-2\gamma
\end{array}\right],
\end{equation}
\begin{equation}
F(t)=\left[\begin{array}{cccc}
i\delta\omega_a(t)&0&0&0\\
0&\!\!\!\!\!\!2i\vartheta(t)-i\delta\omega_a(t)\!\!\!\!\!\!&0&0\\
0&0&i\vartheta(t)&0\\
0&0&0&i\vartheta(t)
\end{array}\right].
\end{equation}
According to the theory of multiplicative stochastic processes the equation
for the ensemble averages can be written as
\begin{equation}
{\rm d\over{\rm d}t}\langle \chi_i\rangle = \sum_j\left( M_{ij}+\sum_k
  Q_{ikkj}\right)\langle\chi_j\rangle,
\end{equation}
where $Q_{ijkl}$ is defined by $\langle F_{ij}(t) F_{kl}(t')\rangle =
2Q_{ijkl}\delta(t-t')$. The quantity $Q_{ijkl}$ has the following form in our
case:
\begin{equation}
Q_{ijkl}=-\delta_{ij}\delta_{kl}\left[\begin{array}{cccc}
\Gamma&-\Gamma&0&0\\
-\Gamma&4L+\Gamma&L&L\\
0&2L&L&L\\
0&2L&L&L
\end{array}\right]_{ik},
\end{equation}
leading to
\begin{equation}
\label{eq:sys1}
{\rm d\over {\rm d}t}\langle\chi_i\rangle = \sum_j N_{ij}\langle\chi_j\rangle
\end{equation}
where
\begin{equation}
N=\left[\begin{array}{cccc}
i\Delta-\Gamma-\gamma\!\!\!\!\!\!\!\!&0&{1\over 2}i\Omega&-{1\over 2}i\Omega\\
0&\!\!\!\!\!\!\!\!-i\Delta-\gamma-4L-\Gamma&-{1\over 2}i\Omega&{1\over 2}i\Omega\\
{1\over 2}i\Omega&-{1\over 2}i\Omega&-L&2\gamma\\
-{1\over 2}i\Omega&{1\over 2}i\Omega&0&\!\!\!-L-2\gamma
\end{array}\right].
\end{equation}
Equations for the ensemble averages of $\chi'_1=\rho_{eg}e^{-i\varphi(t)}$, $
\chi'_2=\rho_{ge}e^{i\varphi(t)}$, $ \chi'_3=\rho_{gg}$, $ \chi'_4=\rho_{ee}$ and
$\chi''_1=\rho_{eg}e^{-2i\varphi(t)}$, $
\chi''_2=\rho_{ge}$, $ \chi''_3=\rho_{gg}e^{-i\varphi(t)}$, $
\chi''_4=\rho_{ee}e^{-i\varphi(t)}$  can be obtained similarly, resulting in
\begin{eqnarray}
\label{eq:sys2}
{\rm d\over {\rm d}t}\langle\chi'_i\rangle &=& \sum_j N'_{ij}\langle\chi'_j\rangle\\
\label{eq:sys3}
{\rm d\over {\rm d}t}\langle\chi''_i\rangle &=& \sum_j
N''_{ij}\langle\chi''_j\rangle,
\end{eqnarray}
where
\begin{equation}
N'=\left[\begin{array}{cccc}
i\Delta-L-\Gamma-\gamma\!\!\!\!\!\!\!\!&0&{1\over 2}i\Omega&-{1\over 2}i\Omega\\
0&\!\!\!\!\!\!\!\!-i\Delta-\gamma-L-\Gamma&-{1\over 2}i\Omega&{1\over 2}i\Omega\\
{1\over 2}i\Omega&-{1\over 2}i\Omega&0&2\gamma\\
-{1\over 2}i\Omega&{1\over 2}i\Omega&0&\!\!\!-2\gamma
\end{array}\right]
\end{equation}
and
\begin{equation}
N''=\left[\begin{array}{cccc}
i\Delta-4L-\Gamma-\gamma\!\!\!\!\!\!\!\!&0&{1\over 2}i\Omega&-{1\over 2}i\Omega\\
0&\!\!\!\!\!\!\!\!-i\Delta-\gamma-\Gamma&-{1\over 2}i\Omega&{1\over 2}i\Omega\\
{1\over 2}i\Omega&-{1\over 2}i\Omega&-L&2\gamma\\
-{1\over 2}i\Omega&{1\over 2}i\Omega&0&\!\!\!-L-2\gamma
\end{array}\right].
\end{equation}
The linear differential equation system composed from
equations (\ref{eq:sys1}), (\ref{eq:sys2}), and (\ref{eq:sys3})  
completely describes the
time evolution of the ensemble 
averages $\langle \rho_{gg}\rangle$, $\langle \rho_{eg}\rangle$, $\langle
\rho_{ge}\rangle$, and $\langle \rho_{ee}\rangle$.

\section{The resonance fluorescence spectrum}
\label{sec:thespectrum}

In the dipole approximation, the resonance fluorescence spectrum
$S(\omega)$ can be calculated as the real part of the two-time
correlation function defined as
\begin{equation}
\Gamma^N_1(\omega)=\lim_{t\rightarrow\infty}\int_0^\infty \exp(-i\omega\tau)\langle\!\langle S^+(t+\tau)S^-(t)\rangle\!\rangle\, d\tau.
\label{correldef}
\end{equation}
Double brackets denote quantum mechanical and stochastical averaging.
The two-time average $\langle\!\langle
S^+(t+\tau)S^-(t)\rangle\!\rangle$ can be expressed by one-time
averages using the quantum regression theorem:
\begin{equation}
\langle\!\langle S^+(t+\tau)S^-(t)\rangle\!\rangle = \left[e^{N''\tau}\right]_{22}\!\!\langle
\chi'_1(t)\rangle + \left[e^{N''\tau}\right]_{24}\!\!\langle \chi'_4(t)\rangle.
\end{equation}
After some calculation one
can obtain the correlation function:
\begin{equation}
\label{eq:spectrum}
\Gamma_1^N(\omega)={\Omega^2\over4\gamma\Omega'}{
    p_1(\omega)p_2(\omega)-\gamma f(\omega)p_2(\omega)+{1\over 2}\Omega^2\over
    p_1(\omega)p_2(\omega)p_3(\omega)+\Omega^2p_4(\omega)},
\label{eq:gamma}
\end{equation}
where
\begin{eqnarray}
\Omega'&=&{\Delta^2\over
  \Gamma+L+\gamma}+\Gamma+L+\gamma+{\Omega^2\over 2\gamma},\nonumber\\
p_1(\omega)&=&i\omega+L+2\gamma,\nonumber\\
p_2(\omega)&=&i\omega-i\Delta+4L+\Gamma+\gamma,\nonumber\\
p_3(\omega)&=&i\omega+i\Delta+\gamma+\Gamma,\nonumber\\
p_4(\omega)&=&i\omega+2L+\Gamma+\gamma,\nonumber\\
f(\omega)&=&\left(1+{i\Delta\over\Gamma+L+\gamma}\right)\left(1-{2\gamma\over
    i\omega+L}\right).\nonumber
\end{eqnarray}

Equation~(\ref{eq:spectrum}), in certain limits, coincides with the
spectra described hitherto. Specifically, when there is no detuning,
the following cases can be distinguished. In the absence of both phase
noise and collisions (when $L \ll \Omega$, $\Gamma \ll \Omega$), it
yields the familiar Mollow-triplet~\cite{Mollow}. In the case of
increasing phase noise and the absence of collisions, the spectrum
will be of a Gaussian shape in the high noise
limit~\cite{KimbleM77}. Increasing collision rate in the absence of
phase noise, the Mollow triplet turns into a doublet structure. In the
high collision rate limit one obtains a spectrum with a narrow dip at
zero frequency that has its origin in quantum
interference~\cite{KarpatiAGLJ02,gawlik}.

The main advantage of the description presented here is that it
enables us to investigate the limit of high collision rate and high
phase noise \emph{simultaneously}. In the case of large phase noise,
there is no dip in the spectrum. Therefore one expects that phase
noise demolishes the quantum interference induced by collisions which
is, according to reference~\cite{KarpatiAGLJ02}, the reason behind the
appearance of the dip.  Our intention is to examine how robust this
interference, introduced in fact by a separate noise process, is
against the competing phase noise.

\begin{center}
\begin{figure}[htbp]
  \includegraphics[angle=-90,width=10 cm]{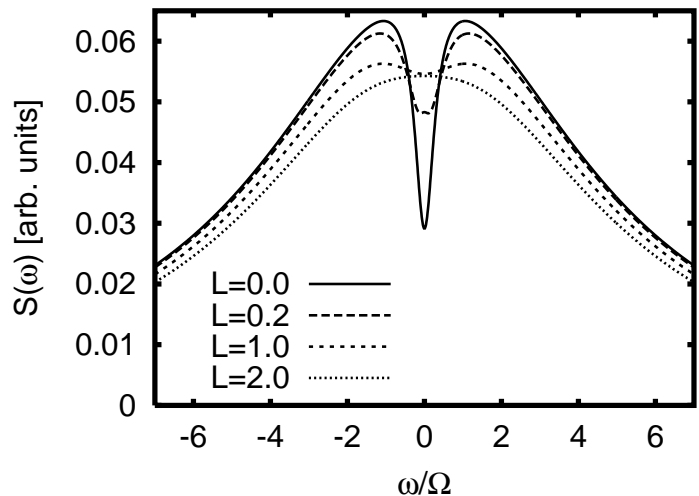}
\caption{
  Resonance fluorescence spectra at parameters $\Gamma/\Omega=5$,
  $\gamma/\Omega=0.05$, $\Delta/\Omega=0$, and at various $L$ parameters: $0$,
  $0.2\Omega$, $1.0\Omega$, $2.0\Omega$.  It can be seen that, though the
  narrow dip in the spectra vanishes as the phase noise increases, it is
  relatively robust against phase noise.}
\label{fig:spectrum}
\end{figure}
\end{center}

The resonance fluorescence spectrum in the case of the high collision
rate case is shown in figure \ref{fig:spectrum} at different phase
noise parameters, illustrating the disappearance of the dip. 

All parameters given in the figures are accessible 
experimentally: $\Gamma$ can be easily adjusted by appropriate buffer 
gas pressure, while $L$ can be also be tailored by manipulating the 
laser cavity, such as, e.g. in \cite{elliott}.

It can be
noticed, that the dip disappears at $L/\Omega\approx 2$, a relatively intense
phase noise. This suggests, that the quantum interference is indeed
robust against phase noise, which will be explained in what follows.

\section{Quantum-trajectory analysis of the interference}
\label{sec:phase}

The stochastically induced quantum interference in coherently driven two-level
system can be analyzed with the aid of quantum-trajectory method. This method
is capable of revealing time correlations between the states of the atom
involved in the interfering transition channels, which are the prerequisites
of quantum interference. Quantum-trajectory methods are based on the
simulation of quantum trajectories, which are individual realizations of the
evolution of the system conditioned on particular sequence of observed events.
By tracking the time evolution of a single quantum-trajectory, the time
correlations can be revealed.
 
An analysis based on quantum-trajectory method was presented for the
phase noise free case in reference~\cite{KarpatiAGLJ02}. A similar
analysis will be carried out here: the quantum-trajectory method of
reference~\cite{DalibardCM92} will be applied for simulating the time
evolution of the system in argument, and phase noise will also be
taken into account.

In this system, a single quantum-trajectory evolves coherently according to
the Hamiltonian
\begin{equation}
\langle H\rangle_{\mathrm{coll}}=\hbar \Delta S^z + {1\over 2}\hbar \Omega(e^{-i\varphi(t)}S^-+S^+e^{i\varphi(t)}),
\end{equation}
interrupted by incoherent gedanken quantum measurements due to collisional
noise events and spontaneous emission represented by the operators
$C_\Gamma=2\sqrt{\Gamma}S^z$ and $C_\gamma=\sqrt{\gamma}S^-$. The simulated
master equation has the form 
\begin{eqnarray}
\dot\rho&=&{1\over i\hbar}[\langle H\rangle_{\mathrm{coll}},\rho] 
-{1\over 2} (C_\Gamma^\dagger C_\Gamma^{\phantom{\dagger}}\rho + \rho C_\Gamma^\dagger
C_\Gamma^{\phantom{\dagger}})+C_\Gamma^{\phantom{\dagger}}\rho
C_\Gamma^\dagger-\nonumber\\
&&-{1\over 2} (C_\gamma^\dagger C_\gamma^{\phantom{\dagger}}\rho + \rho C_\gamma^\dagger
C_\gamma^{\phantom{\dagger}})+C_\gamma^{\phantom{\dagger}}\rho C_\gamma^\dagger,
\label{eq:simmaster}
\end{eqnarray}
that is equivalent to the master equation (\ref{eq:master}). The evolution of
the density operator of the system is obtained by averaging the density
operators of the individual quantum trajectories. The resulting density
operator is the solution of the master equation (\ref{eq:master}).

The accuracy of the simulation is limited by two factors: the length $\Delta
t$ of the time step and  the number $N$ of the simulated quantum
trajectories. $\Delta t$ should 
be much less than the characteristic time of any process in the system. $N$
should be large enough to obtain the right ensemble averages for the density
operator at the given stochastic noise magnitude. In our simulations $N$ was
approximately $5\cdot 10^5$. 

In order to verify the validity of our calculations, we have also derived the
spectra from the numerical results. The results obtained were equal to those
obtainable analytically.

For the further considerations, it is convenient to introduce the
dressed-state basis in which the Hamiltonian in (\ref{eq:hamiltonian}) is
diagonal in the absence of noise:
\begin{eqnarray}
\label{dresseda}
|1\rangle&=&\phantom{-}\cos\Theta |g\rangle + \sin\Theta |e\rangle\\
\label{dressedb}
|2\rangle&=&-\sin\Theta |g\rangle + \cos\Theta |e\rangle
\end{eqnarray}
where
$$
\Theta = -{1\over 2}\arctan\left({\Omega\over\Delta}\right).
$$
An atomic state can be expanded as
$$
|\psi\rangle = c_1 e^{i\varphi_1} |1\rangle + c_2 e^{i\varphi_2}
|2\rangle
$$
in this basis, where $c_1$, $c_2$, $\varphi_1$, $\varphi_2$ are real
numbers. The phase difference and phase sum for the particular state may
then be defined as
\begin{eqnarray}
\label{eq:phdiff}
\Delta\varphi &=&\varphi_1-\varphi_2,\\
\Sigma\varphi &=&\varphi_1+\varphi_2.
\end{eqnarray}
The phase difference can be calculated straightforwardly from a single
quantum-trajectory. It was shown in reference~\cite{KarpatiAGLJ02}, that in
the lack of phase noise, the collisional noise can unexpectedly create phase
correlation between the neighboring atomic dressed states. This is the
underlying physical process that makes the quantum interference possible,
 which  leads to a narrow dip in the spectrum. 
It is then natural to ask
whether phase noise destroys the phase correlation.

\begin{figure}[htbp]
\includegraphics[angle=-90,width=10 cm]{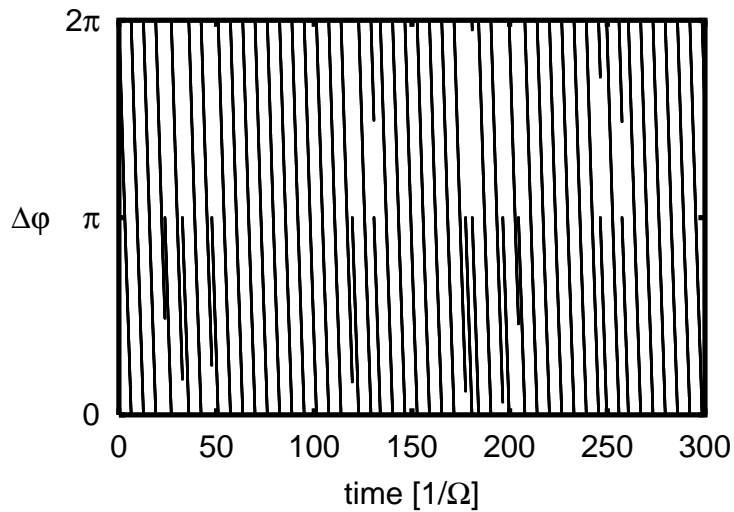}
\caption{The phase difference $\Delta\varphi$ between the dressed states
  for one quantum-trajectory in the case of no phase noise and no
  collisional noise ( $L/\Omega=0$, $\Gamma/\Omega=0$, $\Delta/\Omega=0$,
  $\gamma/\Omega=0.05$). } 
\label{fig:phase1}
\end{figure}

\begin{figure}[htbp]
  \includegraphics[angle=-90,width=10 cm]{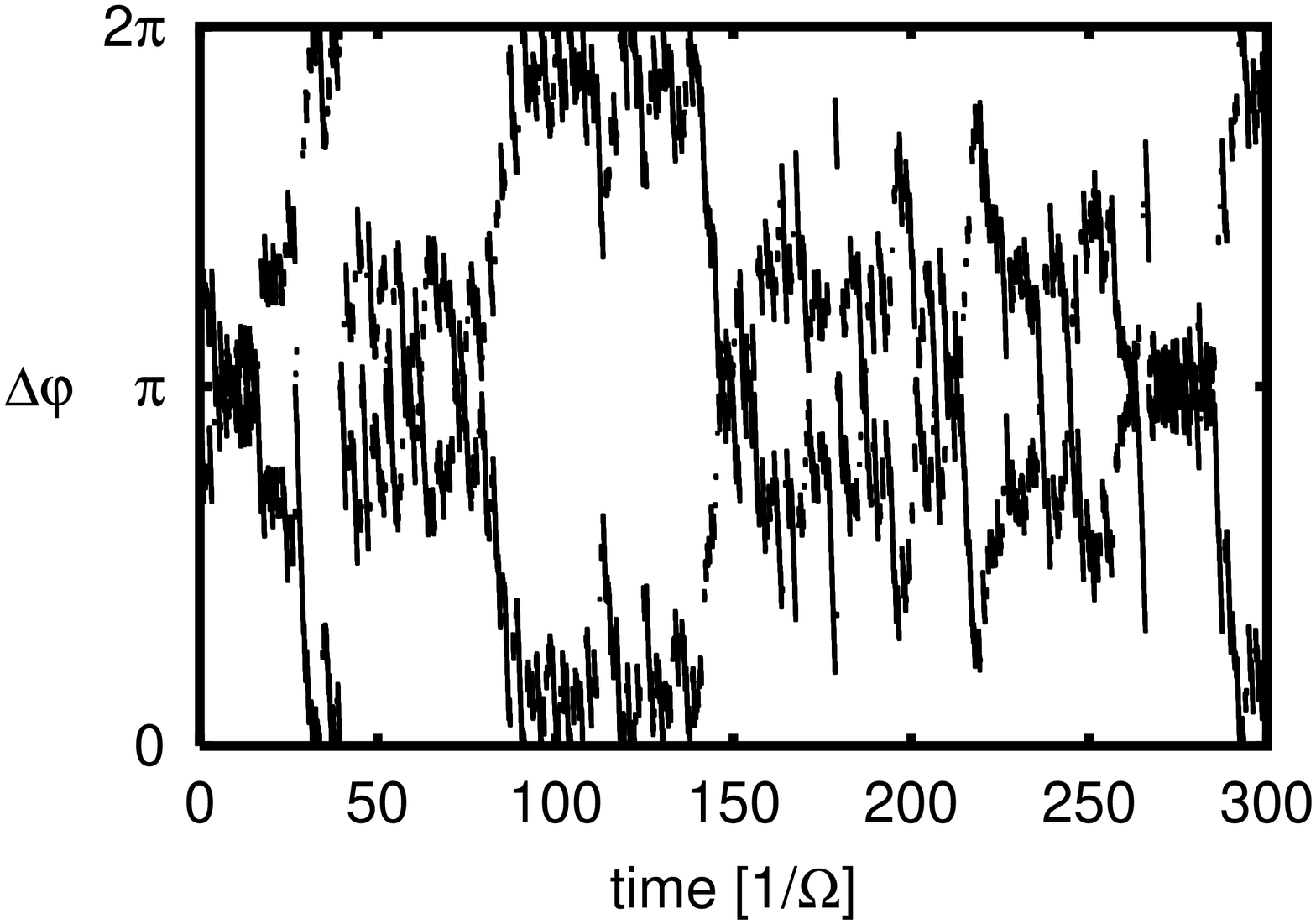}
\caption{The phase difference $\Delta\varphi$ between the dressed states
  for one quantum-trajectory in the case of dominating collisional noise
  ($L/\Omega=0$, $\Gamma/\Omega=5$, $\Delta/\Omega=0$, $\gamma/\Omega=0.05$).  }
\label{fig:phase3}
\end{figure}

\begin{figure}[htbp]
  \includegraphics[angle=-90,width=10 cm]{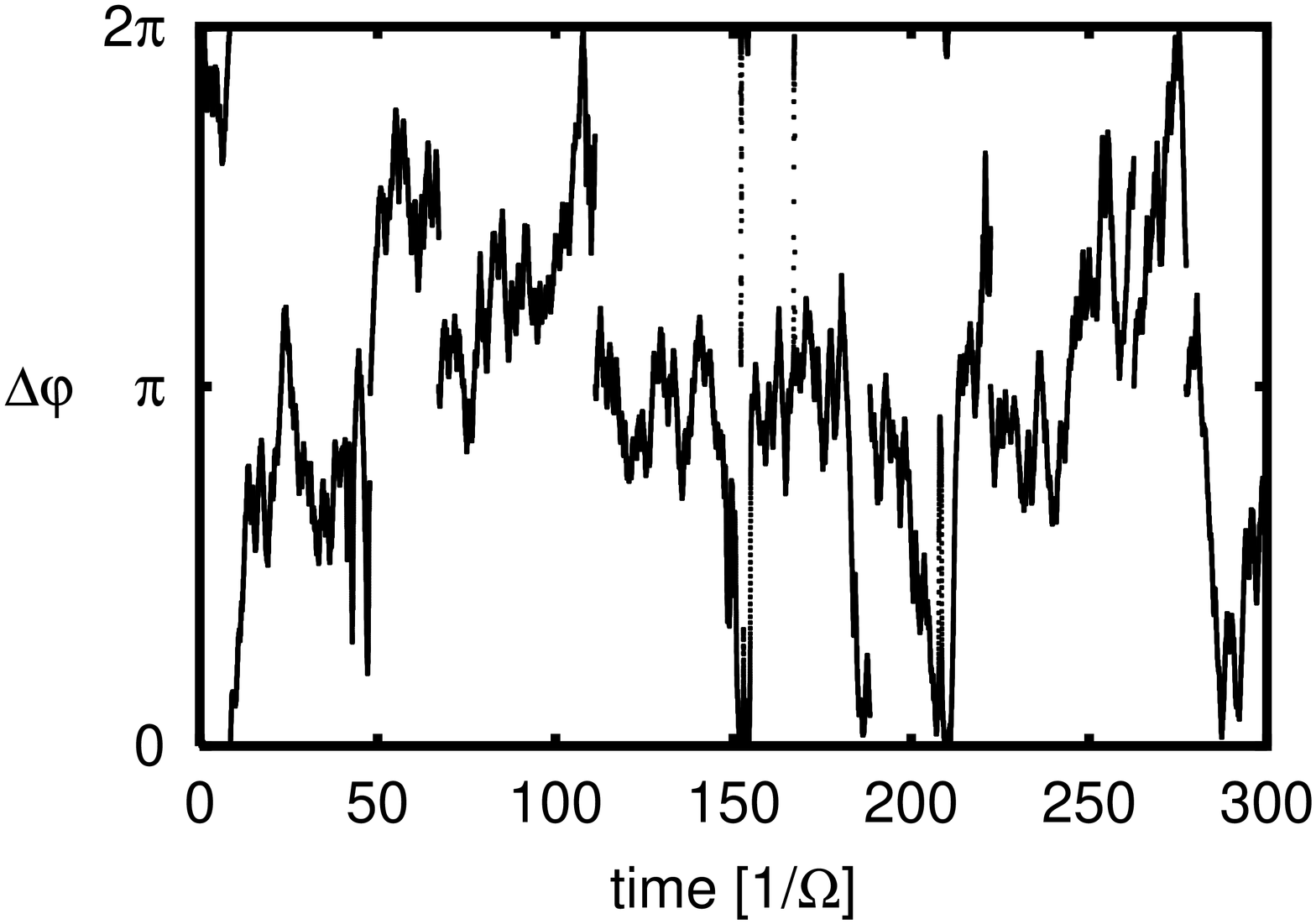}
\caption{The phase difference $\Delta\varphi$ between the dressed states
  for one quantum-trajectory in the case of dominating phase
  noise ($L/\Omega=5$, $\Gamma/\Omega=0$, $\Delta/\Omega=0$,
  $\gamma/\Omega=0.05$). } 
\label{fig:phase2}
\end{figure}

\begin{figure}[htbp]
  \includegraphics[angle=-90,width=10 cm]{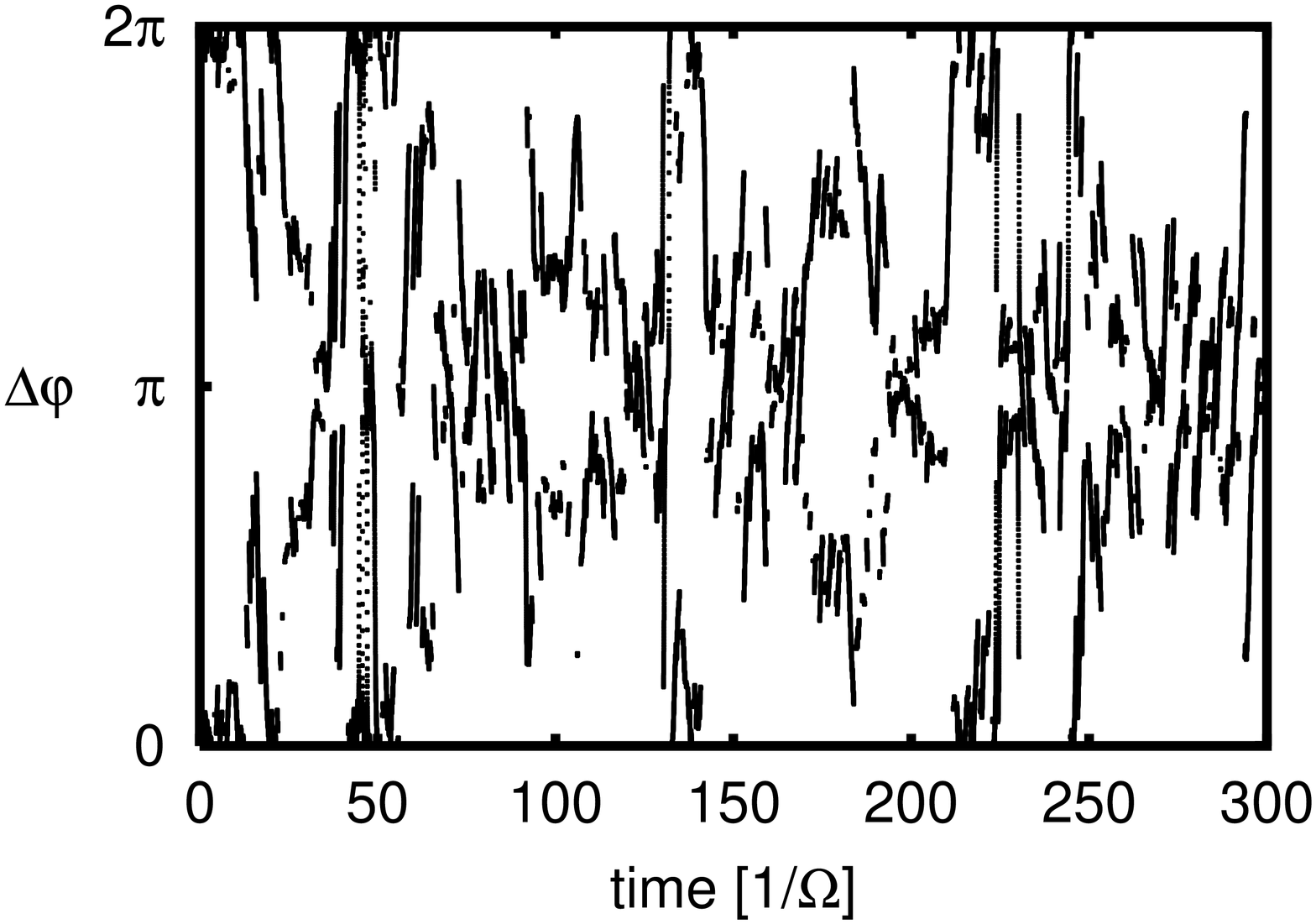}
\caption{The phase difference $\Delta\varphi$ between the dressed states
  for one quantum-trajectory in the case of simultaneous phase
  noise and collisional noise ($L/\Omega=0.5$, $\Gamma/\Omega=5$, $\Delta/\Omega=0$,  $\gamma/\Omega=0.05$).}
\label{fig:phase4}
\end{figure}

The behavior of the phase difference (\ref{eq:phdiff}) for different
settings of the noise parameters are shown in figures
\ref{fig:phase1}-\ref{fig:phase4}. One can observe, that in the lack
of both noise processes (figure~\ref{fig:phase1}) the distribution of
phase difference shows no structure: the time evolution of the phase
difference is linear due to Rabi oscillations, interrupted only by
decays to the ground state due to spontaneous emission.  In the case
of dominant collisional noise (figure~\ref{fig:phase3}), the phase
difference tends to stabilize around the values $0$ and $\pi$. Note
the symmetry of the graph originating from the phase flip caused by
the operator $S^z$ that corresponds to collisional events. The
operator $S^z$ exchanges the two dressed states at resonance
($\Delta=0$): $2S^z|1\rangle = |2\rangle$ and vica versa.  In the high
phase noise without collisional noise (figure~\ref{fig:phase2}), the
phase difference tends to stabilize around the value $\pi$ only. The
most interesting case from our point of view is, however, the fourth
case (figure~\ref{fig:phase4}), when both noises are relevant.  The
structure of the phase distribution is apparently similar to that
without the phase noise.  From this we can conclude, that the phase
correlation created by collisions is indeed not completely destroyed
by the phase noise.

\begin{figure}[htbp]
  \includegraphics[angle=-90,width=10 cm]{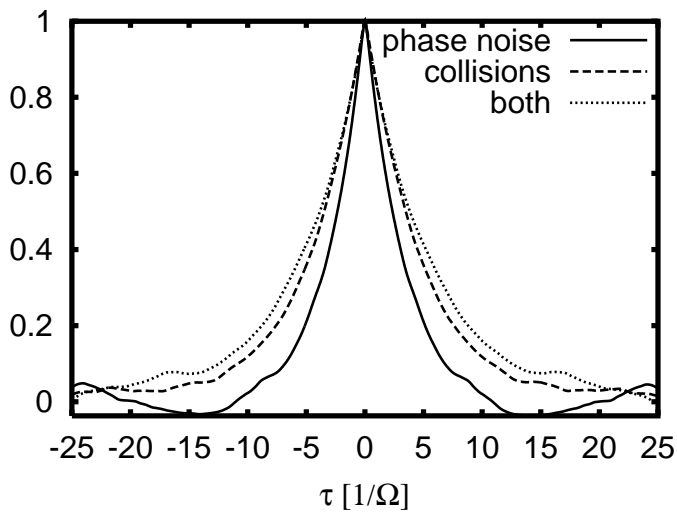}
\caption{The time-correlation function of the cosine of the phase
  difference is shown in the following cases: collisional noise only
  ($L=0$, $\Gamma/\Omega=5$), phase noise only ($L/\Omega=5$, $\Gamma=0$), and both
  noise sources turned on ($L/\Omega=5$, $\Gamma/\Omega=5$). The cosine of the phase
  is correlated in all three cases quite equally. The remaining
  parameters were set to $\Delta=0$, and $\gamma/\Omega=0.05$.}
\label{fig:correl}
\end{figure}

\begin{figure}[htbp]
  \includegraphics[angle=-90,width=10 cm]{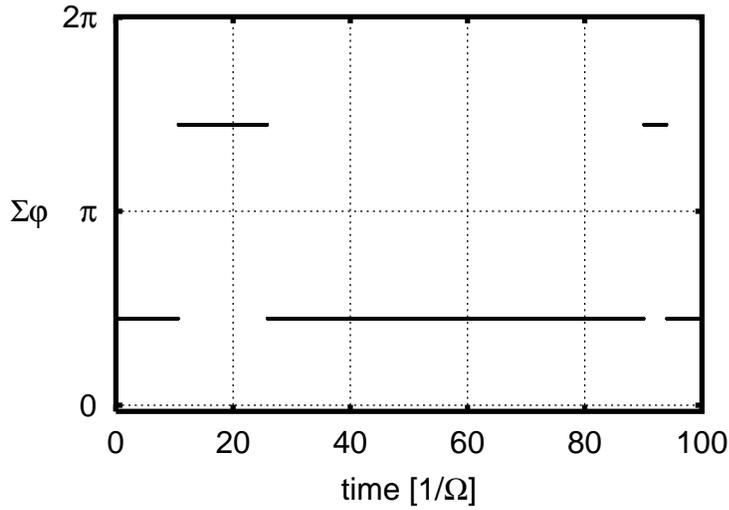}
\caption{The sum of phases in dressed state basis for a single quantum-trajectory. There is no phase noise ($L=0$) in the depicted case,
  the remaining parameters were set to $\Omega=1$, $\Delta=0$, $\Gamma=5$, and
  $\gamma=0.05$. The phase sum remains constant between spontaneous
  emissions.}
\label{fig:nophasenoise}
\end{figure}

\begin{figure}[htbp]
  \includegraphics[angle=-90,width=10 cm]{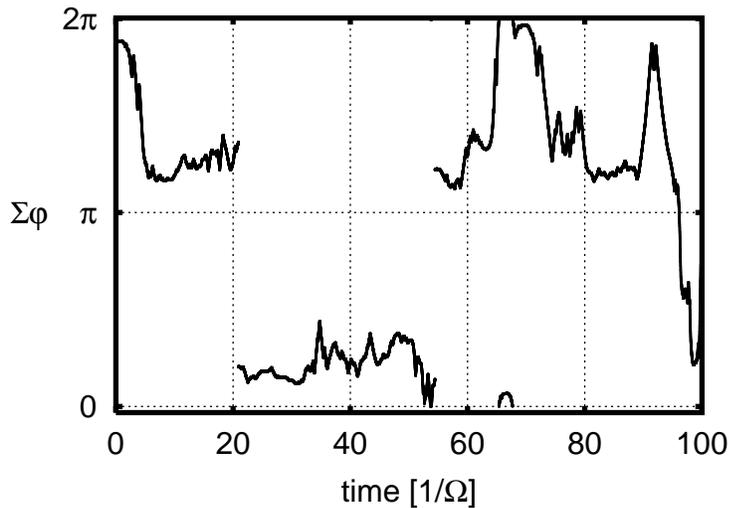}
\caption{The sum of phases in dressed state basis for a single quantum-trajectory. There is a small phase noise ($L/\Omega=0.05$) in the depicted
  case, the remaining parameters were set to $\Delta=0$, $\Gamma/\Omega=5$,
  and $\gamma/\Omega=0.05$. The phase sum changes between spontaneous
  emissions, showing that dressed state mixing occurs due to the phase noise.}
\label{fig:phasenoise}
\end{figure}

Similarly as in Ref.~[19], the phase stabilisation around the $0$ and
$\pi$ phase differences can be characterized explicitly by the
time-correlation function of the cosine of the phase difference,
defined as
\begin{eqnarray}
C_{\cos}(\tau) &=& c \int_{t=0}^T
(\cos\Delta\varphi(t+\tau)-\overline{\cos\Delta\varphi})\nonumber\\*
&&\times (\cos\Delta\varphi(t)-\overline{\cos\Delta\varphi})\, dt,
\end{eqnarray}
depicted in figure \ref{fig:correl}. The figure shows that the phase
noise also leads to phase correlation between the dressed states, moreover,
  the phase stabilisation persist even when both collisional and phase noises
are present in the system.
This on the other hand gives rise to
another question, namely, what is the real reason of the disappearance of
quantum interference.
In order to answer this question, let us transform the Hamiltonian
(\ref{eq:hamiltonian}) at resonance $(\Delta=0)$ to the dressed state basis:
\begin{eqnarray}
H&=&-{1\over2}\hbar \delta \omega_a(t) \left( |1\rangle \langle 2| +
  |2\rangle 
  \langle 1|\right) + 
{1\over 2}\hbar \Omega \bigl( (\cos\varphi(t)
  |1\rangle-i\sin\varphi(t) |2\rangle) \langle  1| 
  + 
\nonumber\\
&&
(\cos\varphi(t) |2\rangle + i\sin\varphi(t) |1\rangle) \langle 2|\bigr).
\end{eqnarray}
It can be seen that the collisional noise $\delta \omega_a(t)$ swaps the
dressed states without disturbing them. Phase noise generates random rotations
in the dressed state basis, mixing the dressed states. This dressed state
mixing can be illustrated by the phase sum of the dressed states during the
time evolution of a single quantum-trajectory. The phase sum remains constant
in collisions that swap the dressed states by the action of $C_\Gamma$. But if
mixing occurs, the phase sum will not remain constant. The phase sum in case
of no phase noise ($L=0$) is depicted in figure~\ref{fig:nophasenoise}, in
case of small phase noise ($L=0.2$) it is shown in
figure~\ref{fig:phasenoise}. The jumps in the figures correspond to
spontaneous emission events.
  
\begin{figure}[htbp]
\includegraphics[angle=0,width=10 cm]{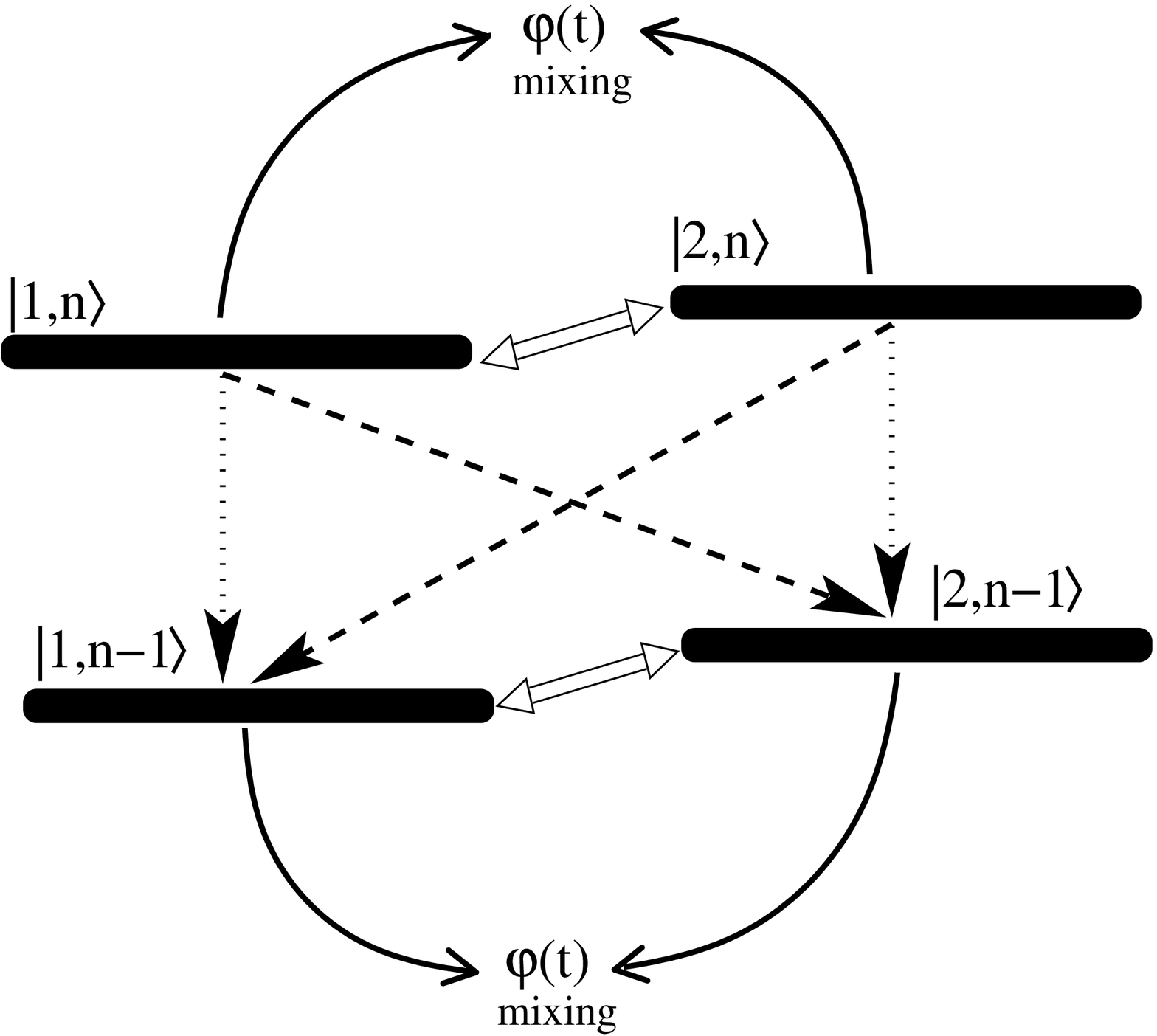}
\caption{Different dressed-state transition channels for a coherently driven
  and stochastically perturbed two-level atom.  The double-stroke arrows
  correspond to the transitions generated by collisional noise. The phase
  noise $\phi(t)$ mixes the dressed state doublets.}
\label{fig:levels}
\end{figure}

Keeping in mind the above considerations, we are now ready to explain the
noise-induced quantum interference phenomenon in the case when both
collisional and phase noises are present. Resonance fluorescence of a
strongly-driven two-level atom is emitted in cascade transitions downward the
ladder of the dressed-state doublets. Figure~\ref{fig:levels} shows two
adjacent doublets and all possible spontaneous and noise-induced transitions
between the dressed-atom states.  Double arrows in
figure~\ref{fig:levels} represent transitions between the dressed states
$|1\rangle$ and $|2\rangle$ generated by the collisional noise events. In the
high noise regime, for some time intervals the phase difference tends to
stabilize not only in the case when either of the noise types are present
separately, but also when simultaneously both of them are present.
If phase noise is relatively low and dressed-state mixing due to phase noise
can be neglected, the two pairs of emission channels $|1,n\rangle\rightarrow
|2,n-1\rangle \rightarrow |1,n-1\rangle$ and
$|1,n\rangle\rightarrow|2,n\rangle \rightarrow |1,n-1\rangle$ (or
$|2,n\rangle\rightarrow |1,n\rangle \rightarrow |2,n-1\rangle$ and
$|2,n\rangle\rightarrow|1,n-1\rangle \rightarrow |2,n-1\rangle$) differ
exclusively by time ordering between collisional mixing and photon emissions.
Photons emitted along these channels are indistinguishable, their destructive
interference results in the dip of the fluorescence spectrum. Increasing phase
noise mixes the dressed states, thus the two channels degenerate to one
channel between two mixed states. Therefore quantum interference disappears.
The mixing of channels, however, becomes relevant only in case of relatively
high level of phase noise, thus the resonance effect is robust against phase
noise.

\section{Summary}
\label{sec:sum}

We have analyzed the quantum interference induced by collisions in a
system of coherently driven two level atoms, in the presence of phase
noise in the driving field. Solving the master equation of the system
in argument analytically, we have derived the resonance fluorescence
spectra. There is a dip in the spectrum when the collisional noise
dominates the Rabi oscillations, in the resonant case. The dip,
introduced by the collisional noise, disappears relatively slowly as
phase noise appears in the driving field.

In order to understand the physical reason behind the relative
robustness of the dip against the phase noise, the underlying quantum
interference phenomena have been investigated. A quantum-trajectory
simulation has been carried out, enabling us to investigate the phase
relations between the dressed states involved in quantum interference
phenomena. We have found that the phase noise does not demolish the
phase correlations introduced by the collisional noise.  Moreover,
phase noise, similarly to the collisional noise, is of a dressed-state
phase-difference stabilizing nature.  On the other hand, it mixes the
dressed-state doublets involved in the quantum interference, which
deteriorates quantum interference in the high-noise limit.

\section*{Acknowledgements}

This work was supported by the Research Fund of Hungary under contract
No. T034484 and by the Polish Committee for Scientific Research
(grant PBZ/KBN/043/PO3/2001). It is also a part of a general program
of the National Laboratory of AMO Physics in Torun, Poland. We thank
Matyas Koniorczyk for the stimulating discussions.

\section*{References}


\begin{thebibliography}{10}

\bibitem{Arimondo96}
Arimondo E 1996 in {\em Progress in Optics XXXV} edited by E.~Wolf (Elsevier,
Amsterdam, 1996) p. 257

\bibitem{ScullyZ98}
Scully M~O and Zhu S~Y 1998 {\em Science} {\bf 281} 1973

\bibitem{CardimonaRS82}
Cardimona D~A, Raymer M~G and Stroud C~R 1982 {\em J. Phys. B-At. Mol. Opt.
  Phys.} {\bf 15} 55

\bibitem{Imamoglu89}
Imamoglu A 1989 {\em Phys. Rev. A} {\bf 40} 2835

\bibitem{ScullyZG89}
Scully M~O, Zhu S~Y and Gavrielides A 1989 {\em Phys. Rev. Lett.} {\bf 62}
  2813

\bibitem{Harris89}
Harris S~E 1989 {\em Phys. Rev. Lett.} {\bf 62} 1033

\bibitem{MandelK93}
Mandel P and Kocharovskaya O 1993 {\em Phys. Rev. A} {\bf 47} 5003

\bibitem{ToorZZ95}
Toor A~H, Zhu S~Y and Zubairy M~S 1995 {\em Phys. Rev. A} {\bf 52} 4803

\bibitem{ZhouS97a}
Zhou P and Swain S 1997 {\em Phys. Rev. Lett.} {\bf 78} 832

\bibitem{ZhuCL95}
Zhu S~Y, Chan R~C~F and Lee C~P 1995 {\em Phys. Rev. A} {\bf 52} 710

\bibitem{ZhuNS95}
Zhu S~Y, Narducci L~M and Scully M~O 1995 {\em Phys. Rev. A} {\bf 52} 4791

\bibitem{XiaYZ96}
Xia H~R, Ye C~Y and Zhu S~Y 1996 {\em Phys. Rev. Lett.} {\bf 77} 1032

\bibitem{ZhuS96}
Zhu S~Y and Scully M~O 1996 {\em Phys. Rev. Lett.} {\bf 76} 388

\bibitem{ZhouS96}
Zhou P and Swain S 1996 {\em Phys. Rev. Lett.} {\bf 77} 3995

\bibitem{ZhouS97}
Zhou P and Swain S 1997 {\em Phys. Rev. A} {\bf 56} 3011

\bibitem{Alzetta}
Alzetta G, Gozzini, Moi L, and Orriols G 1976 {\em Nuovo Cimento} {\bf 36B} 5

\bibitem{PriorBDB81}
Prior Y, Bogdan A~R, Dagenais M and Bloembergen N 1981 {\em Phys. Rev. Lett.}
  {\bf 46} 111

\bibitem{WilsongordonF83}
Wilson-{G}ordon A~D and Friedmann H 1983 {\em Opt. Lett.} {\bf 8} 617

\bibitem{grynberg}
Grynberg G  in {\em Spectral Line Shapes} edited by Exton R (W. De Gruyter,
  Berlin, 1987), Vol.~4, p.\ 503

\bibitem{KarpatiAGLJ02}
Karpati A, Adam P, Gawlik W, Lobodzinski B and Janszky J 2002 {\em Phys. Rev.
  A} {\bf 66} 023821

\bibitem{carmichael-book}
Carmichael H~J {\em An Open System Approach to Quantum Optics} Vol.~M18 of
  {\em Springer Lecture Notes in Physics} (Springer-Verlag, Berlin, 1993)

\bibitem{PlenioK98}
Plenio M~B and Knight P~L 1998 {\em Rev. Mod. Phys.} {\bf 70} 101

\bibitem{GarrawayK94}
Garraway B~M and Knight P~L 1994 {\em Phys. Rev. A} {\bf 49} 1266

\bibitem{DalibardCM92}
Dalibard J, Castin Y and M{\o}lmer K 1992 {\em Phys. Rev. Lett.} {\bf 68} 580

\bibitem{GardinerPZ92}
Gardiner C~W, Parkins A~S and Zoller P 1992 {\em Phys. Rev. A} {\bf 46} 4363

\bibitem{DumZR92}
Dum R, Zoller P and Ritsch H 1992 {\em Phys. Rev. A} {\bf 45} 4879

\bibitem{gawlik}
Gawlik W, {\L}obodzinski B and Cha{\l}upczak W  in {\em Frontiers of Quantum
  Optics and Laser Physics} edited by Zhu S, Scully M and Zubairy M (Springer,
  Berlin, 1997)

\bibitem{elliott}
Elliott D S and Smith S J 1988 {\em J.O.S.Am.B}  {\bf 5} 1927

\bibitem{Agarwal76}
Agarwal G~S 1976 {\em Phys. Rev. Lett.} {\bf 37} 1383

\bibitem{Eberly76}
Eberly J~H 1976 {\em Phys. Rev. Lett.} {\bf 37} 1387

\bibitem{KimbleM77}
Kimble H~J and Mandel L 1977 {\em Phys. Rev. A} {\bf 15} 689

\bibitem{KnightMS78}
Knight P~L, Molander W~A and Stroud C~R 1978 {\em Phys. Rev. A} {\bf 17}
  1547

\bibitem{ToorZ94}
Toor A~H and Zubairy M~S 1994 {\em Phys. Rev. A} {\bf 49} 449

\bibitem{ZhouFZL99}
Zhou P, Fang M~F, Zhou Q~P and Li G~X 1999 {\em Phys. Lett. A} {\bf 251} 199

\bibitem{Agarwal78}
Agarwal G~S 1978 {\em Phys. Rev. A} {\bf 18} 1490

\bibitem{Mollow}
Mollow B~R 1969 {\em Phys. Rev.} {\bf 188} 1969

\end{thebibliography}

\end{document}